\documentclass[aps,superscriptaddress,twoside,twocolumn,floatfix,a4paper,longbibliography]{revtex4-2}

\usepackage{graphicx}
\usepackage{amsmath}
\usepackage{dcolumn}   
\usepackage{bm}        
\usepackage{amssymb}   
\usepackage{graphicx,epsfig}
\usepackage{color}
\usepackage[usenames,dvipsnames]{xcolor}
\usepackage[ocgcolorlinks,colorlinks=true,linkcolor=blue,citecolor=red]{hyperref}
\usepackage{amsmath,amssymb, amsthm, amsfonts}
\usepackage{xspace}
\usepackage{xcolor}
\usepackage{verbatim}
\usepackage{mathtools}
\usepackage{lipsum}
\usepackage{natbib}
\definecolor{goodgreen}{rgb}{0.1,0.5,0}

\usepackage[utf8]{inputenc}
\begin{document}
\title{Phase-control of bipolar thermoelectricity in Josephson tunnel junctions}

\author{Gaia Germanese}
\email{gaia.germanese@phd.unipi.it}
\affiliation{NEST, Istituto Nanoscienze-CNR and Scuola Normale Superiore, I-56127 Pisa, Italy}
\affiliation{Dipartimento di Fisica dell’Università di Pisa, Largo Pontecorvo 3, I-56127 Pisa, Italy}

\author{Federico Paolucci}
\affiliation{NEST, Istituto Nanoscienze-CNR and Scuola Normale Superiore, I-56127 Pisa, Italy}
\author{Giampiero Marchegiani}
\affiliation{Quantum Research Centre, Technology Innovation Institute, Abu Dhabi, UAE}
\author{Alessandro Braggio}
\affiliation{NEST, Istituto Nanoscienze-CNR and Scuola Normale Superiore, I-56127 Pisa, Italy}

\author{Francesco Giazotto}
\email{francesco.giazotto@sns.it}
\affiliation{NEST, Istituto Nanoscienze-CNR and Scuola Normale Superiore, I-56127 Pisa, Italy}

\begin{abstract}
Not so long ago, thermoelectricity in superconductors was believed to be possible  only by breaking explicitly the particle-hole symmetry. Recently, it has been theoretically predicted that a superconducting tunnel junction can develop bipolar thermoelectric phenomena in the presence of a large thermal gradient owing to non-equilibrium spontaneous PH symmetry breaking. The experimental realization of the first thermoelectric Josephson engine then followed. Here, we give a more extended discussion and focus on the impact of the Josephson contribution on thermoelectricity modulating the Cooper pairs transport in a double-loop SQUID. When the Cooper pairs current prevails on the quasiparticle one, the Josephson contribution short-circuits the junction thereby screening the thermoelectric effect.
We demonstrate that the thermoelectric generation due to the pure quasiparticle transport is phase-independent, once Josephson contribution is appropriately removed from the net current measured.
At the same time, we investigate an additional metastable state at $V\approx 0$ determined by the presence of the Josephson coupling, which peculiarly modifies the hysteretic behavior of our thermoelectric engine realized. At the end, we also discuss how the current-voltage characteristics are affected by the presence of multiple thermoelectric elements, which improve the generated output power.
\end{abstract}

\maketitle


\section{Introduction}
After more than a century after the discovery of superconductivity, electronic transport in low-temperature superconductors still represents a very active research field. Superconducting circuits are promising candidates for the realisation of quantum computers~\cite{Arute2019,BlaisRMP93}, and play a crucial role also in fast electronics~\cite{Braginski} and quantum sensing~\cite{Degen2017}. These applications exploit the temperature dependence of the superconducting properties (the kinetic inductance, the resistance, and the superconducting gap $\Delta$), and the nonlinear current-phase relation of the Josephson effect~\cite{barone1982physics}, which is ubiquitous in weak-link structures connecting two superconductors~\cite{LikharevRMP}. In fact, superconducting-based platforms also represent valuable resources for heat management purposes~\cite{GiazottoReview} including on-chip cooling~\cite{Muhonen_2012} and phase-coherent modulation of heat currents~\cite{FornieriReview}. Recently, superconductors have been strongly reconsidered for thermoelectricity~\citep{Virtanen2007,Ozaeta,KolendaPRL,Heikkila2018,Hussein2019,PershogubaPRB99,BlasiPRL,KalenkovJEPT2021,Das2021}, i.e., for the direct conversion of a thermal gradient into an electrical power. At a first glance, thermoelectric effects and the superconducting state are competing for different reasons. First, the thermoelectric currents, associated with single-particle excitations known as Bogoliubov quasiparticles, are typically screened by the dissipationless motion of Cooper pairs. Thus, in bulk superconductors, thermovoltage would be either vanishing or not detectable, as first established in early times~\citep{Meissner, Ginzburg}. Second, a strong contribution to thermoelectric effects is associated with deviations from the particle-hole (PH) symmetry, which typically characterizes normal metals and superconductors~\cite{Ashcroft}. 
Asymmetry between the transport of particles and holes is crucial in the linear response regime, i.e., for voltage $V$ and temperature  bias $\delta T$ which are small with respect to the mean thermal energy. Indeed, a perfect PH symmetry in a two-terminal system, characterized by a fully reciprocal current-voltage (\textit{IV}) characteristics $I(V, \delta T)= -I(-V, \delta T)$~\footnote{For simplicity in the discussion, here we do not include any phase-dependent contributions to the charge current.}, implies a zero charge current $I$ for $V=0$ and, consequently, a zero thermoelectric coefficient in the Onsager matrix. Despite these limitations, superconductors possess some important prerequisites for strong thermoelectric phenomena. One of these ingredients is the strongly energy-dependent quasiparticle density of states (DoS) according to the Bardeen–Cooper–Schrieffer (BCS) theory~\cite{Tinkham2004}. 
This potentiality has been exploited in hybrid superconducting-ferromagnetic insulator tunnel junctions, where the combination of the spin-filtering and the spin-splitting of the BCS superconducting DoS explicitly \textit{breaks} the PH symmetry, leading to strong thermoelectricity~\citep{Ozaeta, BergeretColloquium,KolendaPRL}, thermophase~\cite{GiazottoThermoPhase} effects and nonreciprocal IV characteristics. Other strategies include the identification of non-local effects~\cite{Virtanen2007,Machon,KolendaPRL,KalenkovJEPT2021,Sanchez2018,Hussein2019,Kirsanov2019,BeckmannPRB100,Tan2021} and/or the use of topological materials ~\cite{BlasiPRL,BlasiPRB,Blasi2021,SothmannPRR2,Das2021}.\\
Few years ago, thermoelectricity was theoretically predicted in a superconductor-insulator-superconductor ($S_1IS_2$) tunnel junction with different superconducting zero-temperature energy gaps ($\Delta_{0,1}>\Delta_{0,2}$) and vanishing Josephson coupling ~\citep{MarchegianiPRL,MarchegianiPRB}.
Due to the PH symmetry of superconductors, this thermoelectric effect is intrinsically nonlinear requiring a large temperature gradient above a threshold value. Large temperature gradients determine the condition for a $spontaneous$ breaking of PH symmetry characterising a bipolar thermoelectric effect: both polarities of the generated thermovoltage are possible for a fixed thermal and electronic configuration. In contrast with other systems where opposite signs of the Seebeck voltage are associated to a change of the electronic configuration, in such a system no specific dominant charge carrier is present a priori. Thus, the sign of the thermoelectric signal can be externally selected, or even randomly triggered by noise fluctuations~\cite{MarchegianiAPL}.\\
The nonlinear thermoelectric effect can be also engineered in different structures, such as a tunnel junction composed by a BCS superconductor and a magnetized superconductor $S_m$. As theoretically investigated, the latter can be realized by exploiting the proximity of a superconductor with a ferromagnetic insulator~\cite{GermanesePRB}. In this case, the effect is expected even for identical order parameters ($\Delta_S = \Delta_{S_m}$), since the superconducting energy gap in $S_m$ is effectively reduced by the induced exchange field. Moreover, the thermoelectric response of the junction involves also pure thermo-spin currents due to the spin-split DoS.\\
The proof-of-principle experimental observation of the first bipolar thermoelectric effect in superconducting tunnel junctions was very recently reported~\cite{GermaneseArxiv}. By integrating the tunnel junctions in a Josephson interferometer, an effective and almost-complete suppression of the Josephson coupling is achieved leading to a maximum thermovoltage of $V_{th}\sim 150~\mu$V at sub-Kelvin temperatures. The reported Seebeck coefficient is $S=V_{th}/\delta T\sim 300~\mu$V/K, which is almost the theoretical limit for an aluminium-based structure \cite{MarchegianiPRB}. Notably, it is roughly $10^5$ times larger than the expected value for aluminum in the normal-state at the same temperature. As an engine, this device delivers power of tens of femtowatts to a resistive load kept at room temperature. The demonstrated energy power density of the thermoelectric engine is $\sim$ $140$ nW/mm$^2$, which is a few orders of magnitude smaller than the best theoretical estimate for this device~\cite{MarchegianiPRB}.
Therefore, the best performances of the bipolar thermoelectric engine were reached at the maximum suppression of the critical current, where the Josephson coupling can be neglected.\\
Here, we report a detailed experimental study of the bipolar thermoelectric effect in Josephson junctions focusing on how the Josephson contribution can affect the thermoelectric phenomena. To achieve this aim, we exploit a double-loop superconducting quantum interference device (d-SQUID) to phase-control the thermoelectric performance. In particular, we discuss the interplay of three main contributions: the quasiparticle current, which sustains thermoelectricity, the Cooper pairs tunneling, which is intrinsically reactive, and the Cooper pairs breaking/recombination term. In agreement with the theoretical prediction~\cite{MarchegianiPRR}, we observe that thermoelectricity could survive to the simultaneous presence of the Josephson coupling, until it is completely screened. Furthermore, we develop a thermoelectric engine investigating the metastable state introduced by the Josephson component around $V\approx 0$ that modifies the hysteretic behavior of the device.\\
Besides their relevance in quantum physics, we expect that our results will be pivotal for the development of superconducting electronics~\cite{Braginski}, quantum technologies~\citep{Ladd, Siddiqi, PoliniArxiv}, sensing~\cite{Heikkila2018} and energy harvesting~\citep{Sothmann, Benenti}.\\
The article is organized as follows. We review the essential theoretical models for thermoelectricity in $S_1IS_2$ systems in Sec.~\ref{Sec:Model}. After introducing the device fabrication and the measurement set-up used in Sec.~\ref{Sec:Fab&SetUp}, we discuss the quasiparticle current (Sec.~\ref{Sec:qpEquil}) and the supercurrent contribution (Sec.~\ref{sec:SuperEq}) at thermal equilibrium. 
Then, the resulting thermoelectric effect is characterized by changing the thermal gradient and the temperature bath (Sec.~\ref{Sec:TE}), coming to describe the impact of Josephson coupling on thermoelectricity (Sec.~\ref{Sec:TEJos}). The thermoelectric behaviour of our device is summarized in Sec.~\ref{Sec:TEBath} by changing the injection power, the magnetic flux and the bath temperature. Then, the operation of the device as a heat engine and its obtained performance are shown in Sec.~\ref{Sec:Engine}, discussing as well how Josephson coupling may affect the hysteresis loop of the bipolar thermoelectric engine activated by a current bias. At the end, the implementation of $n$ parallel junctions is theoretically investigated taking into account thermal and structural non-idealities (Sec.~\ref{Sec:Non_idealities}).

\section{Model}
\label{Sec:Model}
In this work, the nonlinear thermoelectricity is based on a standard $S_1IS_2$ Josephson junction. In the tunneling limit, the charge current in a Josephson junction for a direct-current (DC) voltage bias ($V_J$) is generally expressed as~\cite{barone1982physics,Harris1974}
\begin{equation}
I(V_J,\varphi)=I_{\rm qp}(V_J)+I_{j}(V_J)\sin\varphi+I_{\rm int}(V_J)\cos\varphi,
\label{eq:IVphiJJ}
\end{equation}
where $\varphi$ is the phase-difference across the junction. The three terms in Eq.~\ref{eq:IVphiJJ} account for the quasiparticle current ($I_{\rm qp}$), the Cooper pairs tunneling ($I_j$), i.e., the Josephson effect, and the interference contribution due to breaking and recombination processes of Cooper pairs in the two electrodes ($I_{\rm int}$), respectively. In the presence of a voltage bias $V_J$ across the junction, the phase evolves according to the second Josephson equation $\dot\varphi=2eV_J/\hbar$, where $e$ and $\hbar$ are the elementary charge and the reduced Planck's constant, respectively. Thus, the phase-dependent terms of Eq.~\eqref{eq:IVphiJJ} oscillate in time determining a time-dependent voltage $V_J$ due to the circuital electronic components. The result is a complex dynamic time-dependent behaviour, which, for simplicity, can be analyzed resolving a system of coupled nonlinear time-dependent differential equations in the adiabatic approximation ~\cite{MarchegianiPRR}.
This complicated analysis can be conveniently neglected in the DC regime, where the oscillating terms do not affect the charge current average for low values of the critical current.\\
From a thermodynamic perspective, the quasiparticle and the Josephson current are substantially different. Indeed, $I_{qp}$ is associated with thermal excitations resulting in dissipation or power generation, and determining an increasing of the entropy of the system. By contrast, the $I_j$ term, which results finite even at zero-bias [$I_j(V_J=0)=I_c$, with $I_c$ the critical current], represents a purely reactive contribution to the current determined by the equilibrium superconducting ground-state. 
Since the Josephson current is able to short-circuit the junction, it cannot be associated with thermoelectric generation. Instead, the $I_{\rm int}(V_J)$ term can exhibit thermoelectric behavior, but it can be neglected due to the fast time dependence oscillation of the $\cos \varphi$-term, which averages to zero ~\cite{MarchegianiPRR}.\\ 
Therefore, the thermoelectric effects in such systems, characterized in the DC regime, are driven mainly by the first term of Eq.~\eqref{eq:IVphiJJ}. In general, the quasiparticle current can be expressed with the semiconductor modeling ~\cite{Tinkham2004}:
\begin{equation}
I_{qp}=\frac{1}{eR_T}\int_{-\infty}^{+\infty}d\epsilon N_{1}(\epsilon )N_{2}(\epsilon+eV_J)[f_1(\epsilon)-f_2(\epsilon+eV_J)],
\label{eq:iqp}
\end{equation}
where $R_T$ is the normal-state resistance, $N_j$ and $f_j(\epsilon)$ are the quasiparticle normalized DoS and the distribution in the lead $j$ (with $j=1,2$), respectively. For BCS superconductors, $N_j(\epsilon)=|\Re[(\epsilon+i\gamma_j)/\sqrt{(\epsilon+i\gamma_j)^2-\Delta_j^2}]|$, where $\Delta_j$ is the temperature-dependent superconducting energy gap, and $\gamma_j$ is the Dynes parameter accounting for a finite quasiparticle lifetime~\cite{Dynes} ($\gamma=0^+$ in BCS theory).
In our case, when a thermal gradient is applied across the junction, we work in the quasi-equilibrium regime~\cite{GiazottoReview,Muhonen_2012}, where the distribution functions can be written as $f_j(\epsilon)=f(\epsilon,T_j)=[\exp(\epsilon/k_BT_j)+1]^{-1}$ with $T_j$ the electronic temperature of the $j$ lead. Indeed, the electronic temperatures of the two electrodes can be rather different ($T_1\neq T_2$) and decoupled from the lattice phonons temperature ($T_b$), due to the weak electron-phonon interaction at low temperatures and the superconducting nature of the leads.\\ 
The quasiparticle transport can generate thermoelectric power $P=-I_{\rm qp}(V_J)V_J>0$ in the nonlinear regime; yet, linear effects are vanishing due to the leadPH symmetry $N_j(\epsilon)=N_j(-\epsilon)$, resulting in $I_{\rm qp}(V_J)=-I_{\rm qp}(-V_J)$, antisymmetric (reciprocal) in $V_J$ \cite{GermaneseArxiv}. Moreover, the reciprocity in such a system leads to the $bipolarity$ of the thermoelectric effect.\\
The sufficient conditions for this non linear thermoelectricity are the asymmetry between the two zero-temperature superconducting energy gaps ($\Delta_{0,1}>\Delta_{0,2}$)
and a suitable thermal gradient $T_1 \gtrsim T_{2}\Delta_{0,1}/\Delta_{0,2}$, where the superconductor with the larger gap is the hot electrode ($T_1>T_2$) \cite{MarchegianiPRL}. Since the reported effect is non linear, both temperatures have to respect precise threshold conditions.
When all these requirements are fulfilled, the combination of the gapped DoS in the hot electrode ($S_1$) and the monotonically decreasing DoS in energy for $\epsilon\gtrsim \Delta_{0,1}$ in the cold one ($S_2$) results in a thermoelectric current flowing in opposite direction with respect to the applied voltage bias~\cite{MarchegianiPRL}.
Moreover, the presence of thermoelectricity is highlighted by the Absolute Negative Conductance (ANC) [$I_{\rm qp}(V_J)/V_J<0$] in the low bias-regime ($eV_J\ll \Delta_{0,1}+\Delta_{0,2}$). The thermoelectric power is typically maximum for $V_J=V_p(T_1,T_2)=[\Delta_{1}(T_1)-\Delta_{2}(T_2)]/e$, i.e. when the singularities in the DoS of the two electrodes are aligned, and called in the jargon "singularity-matching peak". Finally, the best thermoelectric performance is expected to occur when the gaps ratio is  $r=\Delta_{0,2}/\Delta_{0,1} \sim$ 0.2-0.5 \cite{MarchegianiPRL}.
\section{D-SQUID fabrication and measurement set-up}
\label{Sec:Fab&SetUp}
\begin{figure}[t!]
    \centering
    \includegraphics[width=1\columnwidth]{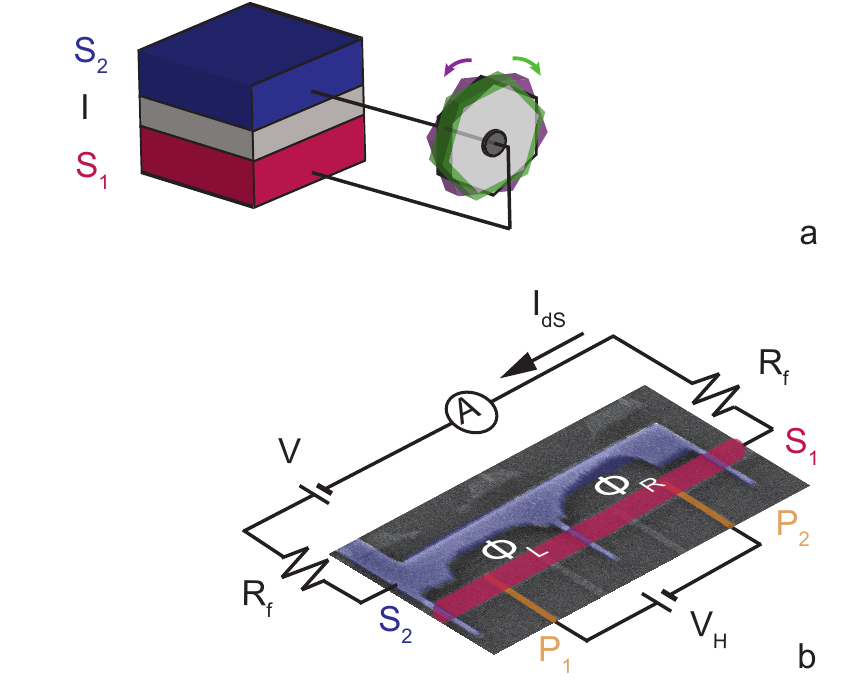}
    \caption{a) A schematic cartoon of the bipolar thermoelectric heat engine is shown, where $S_1$ is the hot Al superconducting electrode (red block), $S_2$ is the cold Al/Cu superconducting bilayer (blue block), and $I$ is the insulating oxide barrier (gray block).
    The thermally-biased $S_1IS_2$ structure is able to generate thermopower in both polarity directions thus leading to the possibility for a gearwheel to rotate clockwise or counter-clockwise (green and violet arrows) for the same temperature gradient (\textit{bipolar} thermoelectricity). b) Experimental measurement set-up. The $S_1IS_2$ d-SQUID in series with two $R_f$ filters is supplied by a voltage ($V$), and the pre-amplified ($A$) current $I_{dS}$ is then measured via a multi-meter. The thermal gradient across the structure is generated by injecting a power $P_{in}$ directly in $S_1$ through a bias voltage ($V_H$) applied to a pair of superconducting tunnel leads ($P_{1,2}$). $\Phi_L$ and $\Phi_R$ are the magnetic fluxes piercing the two d-SQUID rings.}
    \label{Fig1}
\end{figure}

In this section, we introduce the d-SQUID based on three parallel superconducting $S_1IS_2$ junctions, exploited as a bipolar thermoelectric engine in the presence of a fixed thermal gradient (Fig.\ref{Fig1}a).\\
Our device is composed of an aluminum/copper (Al/Cu) bilayer ($S_2$, violet area), which overlaps in three points with the central aluminum strip ($S_{1}$, red area), separated by an Al oxide insulating layer ($I$). Meanwhile, two Al tunnel electrodes ($P_{1,2}$) lie under the Al island (orange area) (Fig. \ref{Fig1}b). The two superconducting rings, formed by $S_1$, $I$, and $S_2$, constitute the d-SQUID, while $P_{1,2}$ act as local electron heaters.\\
The device was nano-fabricated  by a single electron beam lithography (EBL) step, a three-angle shadow mask metals deposition onto a Si wafer, covered with 300-nm-thick thermally-grown $\rm{SiO_2}$, through a suspended bilayer resist mask, and in-situ metal oxidation to create the tunnel junctions. The evaporation and oxidation processes were performed in an ultra-high vacuum (UHV) electron-beam evaporator with a base pressure of $10^{-11}$ Torr. During the evaporation, at first, a 12-nm-thick aluminum film was deposited for $P_{1,2}$ at an angle of $30^\circ$ and, then, exposed to 1 Torr of $O_2$ for 20 minutes to create the $\rm{AlO_x}$ insulating layer (I). Successively, a 14-nm-thick aluminum island ($S_1$) was deposited at $0^\circ$ and oxidized in 1 Torr of pure oxygen atmosphere for 30 minutes to realize the tunnel junctions of the d-SQUID. In the end, a 14-nm-thick aluminum film and 11-nm-thick copper film were deposited at an angle of 30$^\circ$ to form $S_2$ composing the rest of the interferometer.
Indeed, due to the inverse proximity effect~\cite{McMillan, FominovFeigelmanPRB63, MartinisBilayers, BrammertzAPL80}, the thicknesses of the two layers are crucial to determine the superconducting zero-temperature energy gap of $S_2$. In particular, a higher thickness of Copper reduces the zero-temperature energy gap \cite{CatelaniPRB97}. Thus, the AlCu bilayer allows to obtain $\Delta_{0,1}>\Delta_{0,2}$.\\
Figure~\ref{Fig1}b reports a schematic electrical circuit of the provided $I_{dS}V$ measurements with a colored sketched scanning electron microscope (SEM) image.
The measurements were performed in a filtered He$^3$-He$^4$ dry dilution refrigerator (Triton 200, Oxford Instruments) at different bath temperatures ranging from 30 to 650 mK.
The two-wire transport measurements were performed by applying a voltage bias ($V$) across the SQUID through a floating source (GS200, Yokogawa) and recording the current ($I_{dS}$) with a room-temperature current pre-amplifier (Model 1211, DL Instruments). A thermal gradient between the two superconductors was established by the power injected ($P_{in}$) through the two heaters by a battery-powered voltage source (SIM 928, Stanford Research Systems) in the range $V_H=$ 0$-$6.5 mV. Moreover, the magnetic flux through the two loops ($\Phi=\Phi_L+\Phi_R$) of the d-SQUID was provided by a superconducting solenoid driven by a low-noise current source (GS200, Yokogawa).\\
Below, we report the transport measurements as a function of the voltage bias applied across the device ($V=V_J+2I_{dS}R_f$), which corresponds to the sum of the voltage drop across the d-SQUID ($V_J$) and the two RC filters present on the lines of the dilution refrigerator ($2I_{dS}R_f$). As a consequence, the DC Josephson effect appears as a current peak for small but nonzero voltage $V$ (see Sec.~\ref{Sec:qpEquil}). The relation between $V$ and $V_J$ has been theoretically included in the analysis of the experimental curves presented below. This correction is negligible in the subgap region for the characteristics of the superconductors composing the d-SQUID.

\section{Charge transport at thermal equilibrium}
\begin{figure}[ht!]
    \centering
    \includegraphics[width=1 \columnwidth]{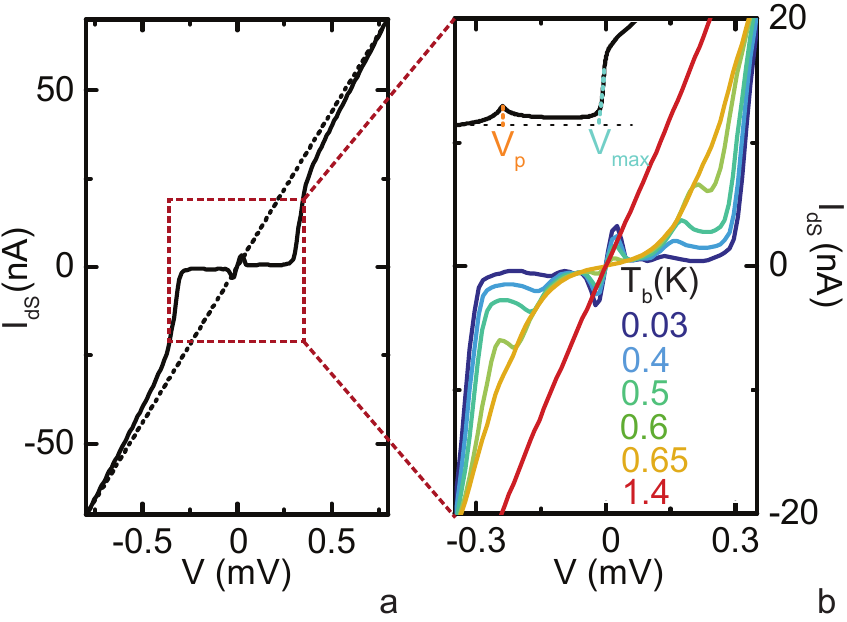}
    \caption{Charge current in the d-SQUID at thermal equilibrium ($P_{in}=$ 0) and zero magnetic flux ($\Phi_L=\Phi_R=0$). a) The current-voltage characteristic at the base temperature of the cryostat ($T_b=$ 30 mK) is displayed. The black dashed line highlights the Ohmic regime occurring for $V\gtrsim 0.6$~mV. The dashed red square indicates the sub-gap region displayed in panel b, where the Josephson current appears around $V=0$. b) The subgap $I_{dS}V$ characteristics are reported for different values of the bath temperature ($T_b$). In the inset, we highlight the quasiparticle peak at $V_p= [\Delta_{0,1}-\Delta_{0,2}]/e$, due to the matching of the singularities of the two superconducting DoSs, and the gap-treshold value $V_{max}= [\Delta_{0,1}+\Delta_{0,2}]/e$.}
    \label{Fig2}
\end{figure}
In this section, we analyse the charge transport in the d-SQUID at thermal equilibrium ($P_{in}=$0). In this case, the electronic temperatures of the two superconductors are equal to the bath temperature ($T_1=T_2=T_b$). In the following subsections, we separately investigate the temperature dependence of the current-voltage ($I_{dS}V$) characteristics of the device, and the phase dependence of the d-SQUID focusing on the Josephson contribution.
\subsection{Characterization of the d-SQUID parameters}
\label{Sec:qpEquil}
Figure~\ref{Fig2}a reports the measured $I_{dS}V$ characteristic at base temperature ($T_b\sim 30~$mK) for zero magnetic flux ($\Phi=0$). The device shows the typical current-voltage evolution of a superconducting tunnel junction~\cite{Tinkham2004}. Indeed, the charge current is strongly suppressed in the subgap region $V<V_{max}$, where $V_{\rm max}= (\Delta_{0,1}+\Delta_{0,2})/e \sim300~\mu$V, while for $V\gg V_{\rm max}$, the d-SQUID shows Ohmic behavior, $I_{dS}\approx V/R_{
tot}$ (black dashed line in Fig.~\ref{Fig2}a). The measured total resistance ($R_{tot}=$ 11 k$\Omega$) includes the series of two RC filters in the cryostat measurement lines ($R_f$= 1.1 k$\Omega$) and the d-SQUID normal-state resistance ($R_T=$ 8.8 k$\Omega$), resulting from the parallel of the three tunnel junctions. The DC Josephson current appears as a peak around $V\approx0$ due to the two-wire measurement setup. \\
The sub-gap current is displayed in Fig.~\ref{Fig2}b for different values of $T_b$. By increasing $T_b$, the supercurrent is reduced, while the quasiparticle current strongly rises due to thermal excitation. For $T_b\gtrsim 0.4~$K, the quasiparticle current clearly shows a non-monotonic dependence in the voltage bias over the Josephson peak, highlighted by the rise of the singularity-matching peak $V_p$, as defined in the previous section (see inset of Fig.~\ref{Fig2}b). In this way, the zero-temperature values of the superconducting energy gaps can be estimated from $V_p(T_b)$ and $V_{max}(T_b)$~\footnote{A more precise determination is obtained through numerical fitting of the equilibrium IV characteristic, as discussed in in the Supplementary of Ref.~\cite{GermaneseArxiv}}, obtaining $\Delta_{0,1}\approx 220 ~\mu$eV and $\Delta_{0,2}\approx 80 ~\mu$eV. The optimal values for the Dynes parameters can be extracted from the same fitting ($\gamma_1 =2\times10^{-3} \Delta_{0,1}$ and $\gamma_2 =2\times10^{-2} \Delta_{0,2}$, see Ref. \cite{GermaneseArxiv} for details).  
Thus, the energy gaps ratio is $r\approx 0.36$, which is within the range of the best thermoelectric performance expected for this device \cite{MarchegianiPRL}.\\
The evolution of the charge current as a function of the temperature allows estimating the critical temperature ($T_{c,j}$) of the $S_j$ superconductor (with $j=1,2$). In particular, $T_{c,1} \sim 1.4~$K is evaluated by considering the complete reduction of $V_{max}$, when even $S_1$ switches to the Ohmic regime (red linear dependence in Fig.~\ref{Fig2}b). Instead, the estimation of the critical temperature of the bilayer ($T_{c,2} \sim$ 650 mK) can be deduced from the evolution of the Josephson current peak around $V\approx 0$, which is suppressed with increasing $T_b$~\cite{Ambegaokar} (orange curve in Fig.~\ref{Fig2}b).
It should be noticed that the ratio $\Delta_{0,2}/(k_B T_{c,2})$ slightly deviates from the standard value predicted by the BCS theory ($\alpha_{BCS}= \pi/e^{\gamma_E}\sim$ 1.764, where $\gamma_E$ is the Euler's constant). This behavior is a probable consequence of the superconductor-normal metal bilayer nature of $S_2$ (see Supplementary information of Ref.~\cite{GermaneseArxiv}).\\
The energy gap and the critical temperature of the superconducting probes ($P_{1,2}$), used as heaters, are extrapolated from the current-voltage characteristic of $P_1IS_1IP_2$ junctions, and present   values similar to those of $S_1$: $\Delta_{0,P}\simeq 220 \mu$eV (not shown here). The total resistance of the series connection of the two tunnel electrodes is $R_{T,P}=$ 51 k$\Omega$. The sum of the gaps involved into the $I_{dS}V$ characteristic of $P_1IS_1IP_2$ junction [$V_{max}= 2(\Delta_{0,P}+\Delta_{0,S_1})/e= 0.88$ mV] determines the minimum voltage bias to be applied across the tunnel probes to ensure heating of $S_1$, corresponding to an injected power $P_{in} \gtrsim V^2_{max}/2R_{T,P}= $10 pW.\\

\subsection{Phase-dependent behaviour of the d-SQUID}
\label{sec:SuperEq}
\begin{figure}[ht!!]
    \centering
    \includegraphics[width=1 \columnwidth]{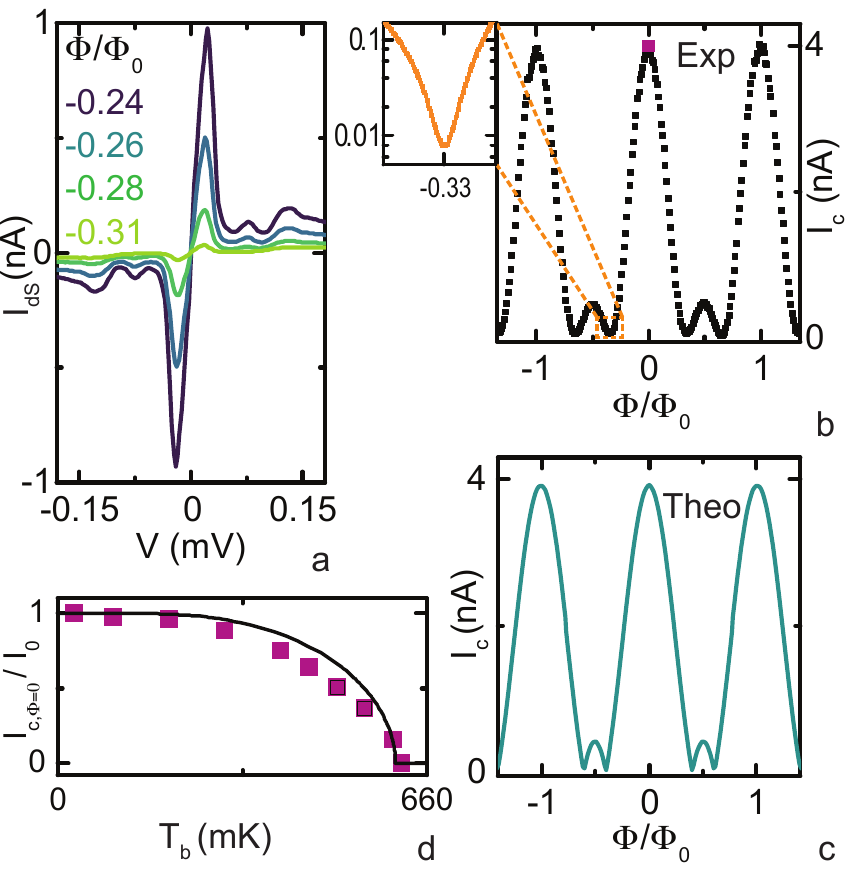}
    \caption{Flux-modulation of the Josephson current at thermal equilibrium ($P_{in}=0$). a) The current-voltage evolution close to zero-bias for different values of the reduced magnetic flux is reported. The critical current $I_c$ is modulated by the flux. b) The critical current is displayed as a function of the reduced flux $\Phi/\Phi_0$, expressing the periodic interference pattern of the d-SQUID. Little bumps are present between two maximum peaks. A blow-up of one minimum is shown in the left inset reaching $I_c=$7 pA. c) The theoretical model of the d-SQUID interference pattern is shown at base temperature. It is in good agreement with the experimental data. d) The critical current normalized is shown as a function of the bath temperature. The behavior follows the Ambegaokar-Baratoff relation (solid line).}
    \label{Fig3}
\end{figure}
The Josephson current flowing across the d-SQUID can be modulated by varying the magnetic flux piercing the loops. In Fig.~\ref{Fig3}a, the $I_{dS}V$ characteristics are shown for different values of $\Phi/\Phi_0$ (with $\Phi_0$ the magnetic flux quantum) leading to the supercurrent modulation around the zero-voltage. The maximum of the Josephson current is deduced from the $I_{dS}V$ characteristics for each value of the magnetic flux and reported in the d-SQUID interference pattern of Fig.~\ref{Fig3}b. The minimum critical current value reached [$I_{c}(\Phi/\Phi_0=-0.33)\sim$ 7 pA, see the inset of Fig.~\ref{Fig3}b] is 1.75\textperthousand ~of its maximum [$I_{c}(\Phi=0)=4$~nA, purple square]. This notable result helps to highlight the quasiparticle contribution guaranteeing a clear manifestation of thermoelectricity in the presence of negligible Josephson effect.
~\cite{MarchegianiPRR}.\\ 
The double-loop interferometer, formed
by three Josephson junctions, is well described by the following model, which assumes a sinusoidal current-phase relation and considers the fluxoid quantization constraint. The critical current $I_c$ is a function of the magnetic fluxes through the left and right loops ($\Phi_L$, $\Phi_R$). Assuming a negligible inductance of the rings and maximizing with respect to the phase difference in the central Josephson junction, the relation can be written as a function of $\Phi/\Phi_0$~\cite{FornieriDoubleLoop}:
\begin{align}
    \left[\frac{I_c(\Phi/\Phi_0)}{I_0}\right]^2=
&1+\sum_{j=\pm}{r_j}^2+2r_j \cos[2\pi(1+j\alpha)\Phi/\Phi_0]
\nonumber
\\
    &+2r_+ r_- \cos(2\pi\Phi/\Phi_0),
\end{align}

where $I_0$ is the critical current of the central junction, $r_+=I_L/I_0$ and $r_-=I_R/I_0$ are the normalized critical currents of the
lateral junctions. The effective area difference between right/left loop geometry determines the flux-asymmetry coefficient, that is $\alpha=(\Phi_L-\Phi_R)/(\Phi_L+\Phi_R)$. \\
The theoretical curve shown in Fig.~\ref{Fig3}c is in good agreement with the experimental data, allowing us to deduce precise device parameters. Indeed, $r_+=0.63$ and $r_-=0.65$ satisfy the triangle inequality $|r_+-r_-|\leq 1$ and $r_++r_- \geq 1$, justifying an excellent suppression of the supercurrent. In this way, the double-loop interferometer provides a more robust suppression of the supercurrent with respect to the common approach with the single-loop SQUID~\cite{KemppinenAPL92,RonzaniAPL104,GranataPhysRep}, even though the loops are not exactly identical. As shown in the SEM image of Fig.\ref{Fig1}b, the central junction area is larger than the lateral ones leading it to be less resistive ($R_L=$ 31.8 k$\Omega$, $R_0=$ 20 k$\Omega$, $R_R=$ 30.7 k$\Omega$). In this way, the critical current, which is inversely proportional to the tunnel resistance \citep{LikharevRMP, Golubov}, is higher in the central lead.
Moreover, when $I_i>0.5I_0$ (with i$=$ L, R),  additional small bumps appear in the periodic interference pattern. The flux-asymmetry coefficient is low ($\alpha=0.003$), since the two loops are designed with the same nominal area ($A_1=A_2 \simeq 1.6~\mu {\rm m}^2$, see the SEM picture of Fig.\ref{Fig1}b).\\
The critical current normalized with respect to the maximum value reached $I_{c}/I_0$ decreases monotonically with the bath temperature, as shown in Fig.~\ref{Fig3}d. In particular, we observe that the critical current completely disappears around $T_b=$ 650 mK, which is the critical temperature value of the bilayer $S_2$, and follows the Ambegaorkar-Baratoff relation \cite{Ambegaokar}, reported in solid line.

\section{Thermoelectric effect}
\label{Sec:TE}
\begin{figure}[ht!]
    \centering
    \includegraphics[width=1 \columnwidth]{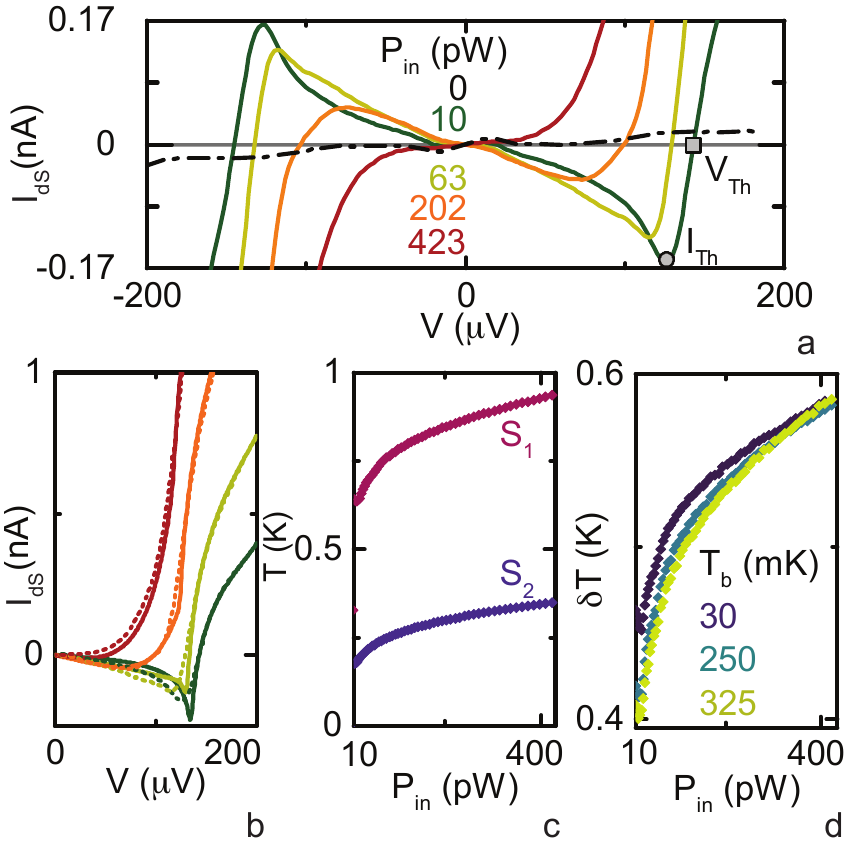}
    \caption{Charge current and thermoelectricity with power injection ($P_{\rm in}\neq 0$) at the minimum Josephson coupling $\Phi=-0.33\Phi_0$. a) The subgap $I_{dS}V$ characteristics are shown for different values of $P_{\rm in}$ (colored solid lines). For comparison, the $I_{dS}V$ characteristic for $P_{in}=$ 0 is reported in black dashed line. The maximum thermoelectric current $I_{Th}$, and the Seebeck voltage $V_{Th}$ for $P_{in}=10~$pW are marked with a circle and a square, respectively. b) The theoretical $I_{dS}V$ characteristics (solid lines) limited to the subgap regime are shown in comparison with the experimental data (dashed lines) for different values of the input power at base temperature. c) Electronic temperature calibration at $T_b=30~$mK of the two leads (red for $S_1$,  and blue for $S_2$) is displayed as a function of the input heating power, computed through fitting of the current-voltage characteristics.  
    d) The estimated temperature gradient ($\delta T= T_1-T_2$) is reported as a function of the injected power for different values of $T_b$.}
\label{Fig4}
\end{figure}
\begin{figure}[ht!]
    \centering
    \includegraphics[width=1 \columnwidth]{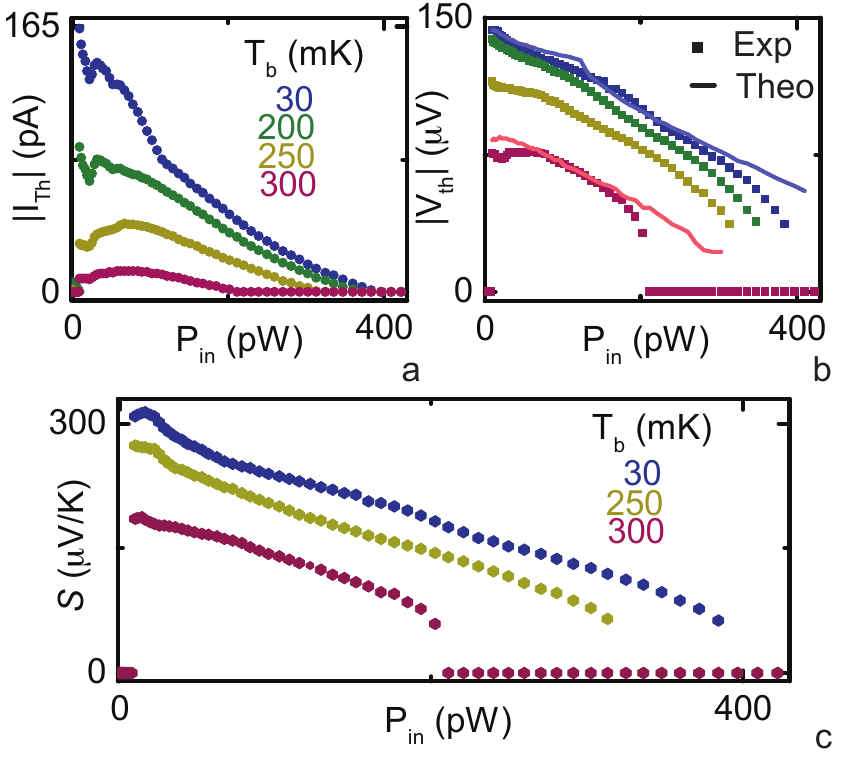}
    \caption{Thermoelectric figures of merit as a function of the input power for different values of the bath temperature. a) The absolute value of the maximum generated thermocurrent.   
    b) The absolute value of the Seebeck voltage.
    Solid lines are the theoretical values extracted from the fitting curves. 
    c) Nonlinear Seebeck coefficient.}
    \label{Fig5}
\end{figure}
In this section, the main results on thermoelectricity are shown and commented. Once fixed the magnetic flux to minimize the critical current ($\Phi/\Phi_0=-0.33$), we discuss the charge current that flows through the d-SQUID by changing $P_{in}$ and $T_b$.\\   
In Fig.~\ref{Fig4}a the experimental $I_{dS}V$ characteristics (solid lines) are shown for $T_b=30~$mK at different values of $P_{in}$ heating $S_1$ through the tunnel probes. This heating power determines an electronic temperature difference between $S_1$ and $S_2$, that is $T_1>T_2$.
For a comparison, we also include the curve measured at $P_{\rm in}=0$ (black dashed-dotted line).\\
In the subgap regime ($V< 300 ~\mu$eV), the standard dissipative behavior [$I_{dS}(V)V>0$] is displayed for $P_{\rm in}=$0. The minimal value of the injected power to activate the nonlinear thermoelectricity is $P_{\rm in}=$ 10 pW,  owing to the superconducting nature of the heaters, whose energy gap determines the minimum voltage bias to be applied across the tunnel probes to induce Joule heating in $S_1$.
Thus, for $P_{in}=$ 10-400 pW an appropriate thermal difference between the two leads is established ($T_1\gtrsim T_2\Delta_{0,1}/\Delta_{0,2}$~\cite{MarchegianiPRB}), and a thermoelectric current is generated against the voltage bias applied [$I_{dS}(V)V<0$]. When the input power is too large ($P_{in}>$ 400 pW), the $I_{dS}V$ characteristics return being dissipative. The reason for this behavior could lie on the consequent heating of the cold lead at higher injected power: once reached the threshold value for the cold lead, the system ceases to be thermoelectric, even though a large thermal gradient is still applied across the junction.\\
Moreover, by increasing $P_{in}$ also the residual critical current is reduced, and the ANC, defined as $I_{dS}(V)/V<0$, appears also around $V\approx 0$ ~\cite{MarchegianiPRB}.\\
The antisymmetric (reciprocal) current obtained in the voltage bias implies as well the bipolarity of the thermoelectric effect. For a fixed thermal gradient, both polarities of the thermovoltage could be generated across the junction. This evidence is a direct consequence of spontaneous PH symmetry breaking, which stands out in our system \cite{MarchegianiPRL}.\\
Thus, the setup we are considering is intrinsically nonlinear. The main thermoelectric figures of merit that identify these curves are the maximum thermoelectric current generated roughly at the matching peak $I_{Th}(V_p)$ (gray circle), and the thermovoltage $V_{Th}$ (gray square), i.e., the voltage drop across the tunnel junction when $I_{dS}(V_{Th})=0$. \\ 
The experimental curves in the subgap regime measured at the maximum suppression of the critical current (dashed lines) are compared with the theoretical ones (solid lines) in Fig. \ref{Fig4}b by using the best fit parameters for four values of $P_{in}$ shown at the base temperature. For simplicity, our theoretical model [Eq.~\eqref{eq:iqp}] takes into account only the quasiparticle current for a single $S_1IS_2$ junction in the subgap region [$I_{qp}(V, T_1, T_2)$, for $V< V_{max}$] neglecting the Josephson contributions. The temperatures of the two leads are the only free parameters of the out of equilibrium fit. Indeed, the tunnel resistance, the temperature-dependent gaps and the Dynes parameters are obtained in the equilibrium analysis ($T_1=T_2=T_b$) (more details are provided in Supplementary Information of Ref.\cite{GermaneseArxiv}). The not-perfect matching between the experimental data and the theoretical fit, especially at the current peak, is probably due to our simplified model, and will be discussed in Sec.\ref{Sec:Non_idealities}.\\
Differently, in the presence of thermoelectricity ($P_{in}=10-400$ pW), the calibration of two leads temperature was extracted from the best fits as a function of the input power [$T_1(P_{in}), T_2(P_{in})$], as reported in Fig. \ref{Fig4}c. The temperature of $S_1$ monotonically increases from 650 mK to 930 mK for values of $P_{in}$ showing thermoelectricity. At the same time, $T_2$ rises from 200 mK to 340 mK. The overheating estimated for the cold lead is reasonable considering all thermal channels, that is the heat-current flowing from the hot reservoir to the cold one and the electron-phonon scattering in $S_2$.\\
Iterating this analysis for all $I_{dS}V$ characteristics at different values of $P_{\rm in}$ and $T_b$, we obtain the effective thermal gradient between the two superconductors ($\delta T= T_1-T_2$), as shown in Fig.~\ref{Fig4}d. We note that the minimum thermal gradient request to generate thermoelectric current is around $450$ mK at $T_b=30$ mK. We observe that $\delta T$ decreases by increasing $T_b$ affecting the temperature conditions for thermoelectricity at low heating powers. This behaviour can be ascribed to a sizeable change of the electron-phonon thermalization in $S_1$ dependent on the bath temperature. Moreover, the cold side starts to be heated by the thermal current, which flows between the two terminals.\\
The evolution of $\delta T$ affects the main thermoelectric properties ($|I_{Th}|$, $|V_{Th}|$), shown in Fig.~\ref{Fig5} as a function of $P_{in}$. Indeed, by increasing $T_b$ the thermoelectricity decreases until its full suppression for $T_b>$ 350 mK. This behaviour is opposite to the standard linear thermoelectricity, where the performance are proportional to the temperature gradient, due to the non-linearity of the system. 
In particular, the thermocurrent generated ($|I_{Th}(V_p)|$) decreases non-linearly with $P_{in}$, as displayed in Fig.~\ref{Fig5}a. The maximum current generated is $162$ pA occurring for $P_{in}=$ 10 pW at $T_b=$ 30 mK. For low bath temperatures ($T_b \lesssim 0.4 T_{c,2} \sim$ 260 mK), where $\Delta_2(T) \simeq \Delta_{0,2}$, the thermoelectric current is maximum at $P_{in}=$ 10 pW, when thermoelectricity starts manifesting. Instead, for higher temperatures ($T_b \geq 0.4 T_{c,2}$), where $\Delta_2(T) < \Delta_{0,2}$, the maximum of $I_{Th}$ is reached for higher input power ($P_{in}=$ 90 pW). 
This behavior may depend on the exponentially-damped electron-phonon relaxation in $S_1$ at low temperatures owing to the presence of the superconducting energy gap \citep{GiazottoReview, HeikkilaReview, Timofeev}.\\
In the same way, the dependence of the thermovoltage $|V_{Th}|$ on $P_{in}$ is displayed in Fig.~\ref{Fig5}b for different values of $T_b$. At $T_b=$ 30 mK, the thermovoltage decreases monotonically for $P_{in}\gtrsim10~$pW reaching a maximum value of $\sim142~\mu$V. This non-linear behavior is in contrast with what expected for the linear thermoelectricity, which generally increases with the thermal gradient, and, therefore, with the injected power ($V_{th}\propto \delta T \propto P_{in}$) \cite{Benenti}. The corresponding theoretical predictions (solid curves in Fig.~\ref{Fig5}b) are computed for two bath temperatures ($T_b=30$~mK and 300 mK) by considering the temperature calibration of the two leads. The curves are in good agreement with the experimental results confirming once again the validity of the quasiparticle transport model used.\\ 
In light of the experimental and theoretical analysis carried out, we report in Fig.~\ref{Fig5}c the nonlinear Seebeck coefficient ($\mathcal{S}=|V_{Th}|/\delta T$) as a function of $P_{in}$ at different bath temperatures, which describes the magnitude of the thermovoltage generated in response to a temperature difference. Thus, $|V_{Th}|$ decreases monotonically with $P_{in}$ while increasing $\delta T$ (see Fig.~\ref{Fig4}d). Similarly, $\mathcal{S}$ is reduced by increasing $T_b$. The maximum value  $\mathcal S\sim$ 308 $\mu$V/K is five orders larger than the Seebeck coefficient for a normal metal at the same temperature~\citep{Mott, Mamin}. Even for $T_b=$ 300 mK, we still find huge nonlinear Seebeck coefficients up to around 160 $\mu$V/K.

\subsection{Impact of the Josephson effect on thermoelectricity} \label{Sec:TEJos}
\begin{figure}[t!]
    \centering
    \includegraphics[width=1 \columnwidth]{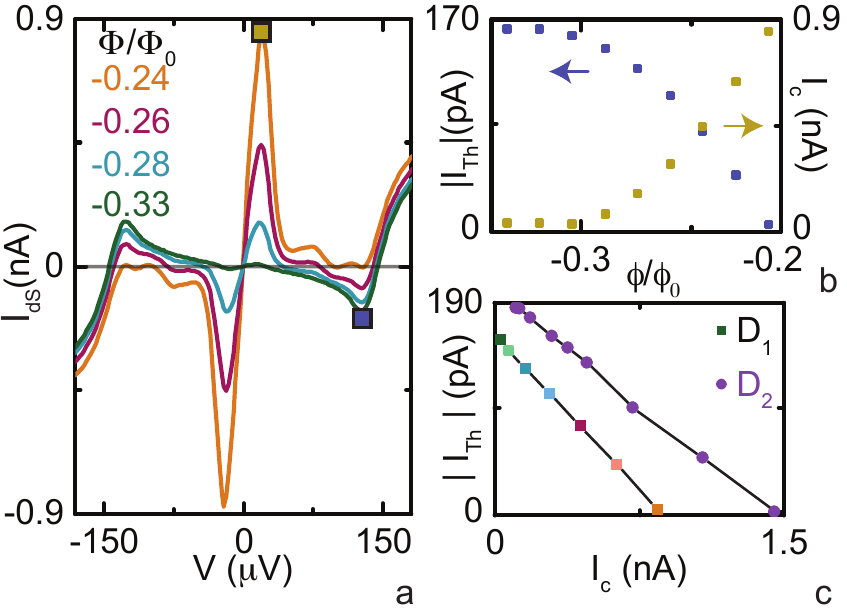}
    \caption{a) $I_{dS}V$ characteristics for different values of $\Phi/\Phi_0$ with $P_{in}=$10 pW at the base temperature. The golden and the blue squares are referred to the maximum of the critical current and the thermoelectric current generated at the peak, respectively. b) The absolute value of the thermoelectric current (blue squares) on the left axis, and the critical current (golden squares) on the right axis are reported as a function of the normalized magnetic flux. c) The maximum absolute value of the thermoelectric current generated $|I_{Th}|$ as a function of the maximum critical current $I_c$ is reported for the device $D_1$  analyzed in this work (colored squares), and another one $D_2$ (violet circles), characterized in the Supplementary Information in \cite{GermaneseArxiv}.}
    \label{Fig6}
\end{figure}

\begin{figure}[t!]
    \centering
    \includegraphics[width=1 \columnwidth]{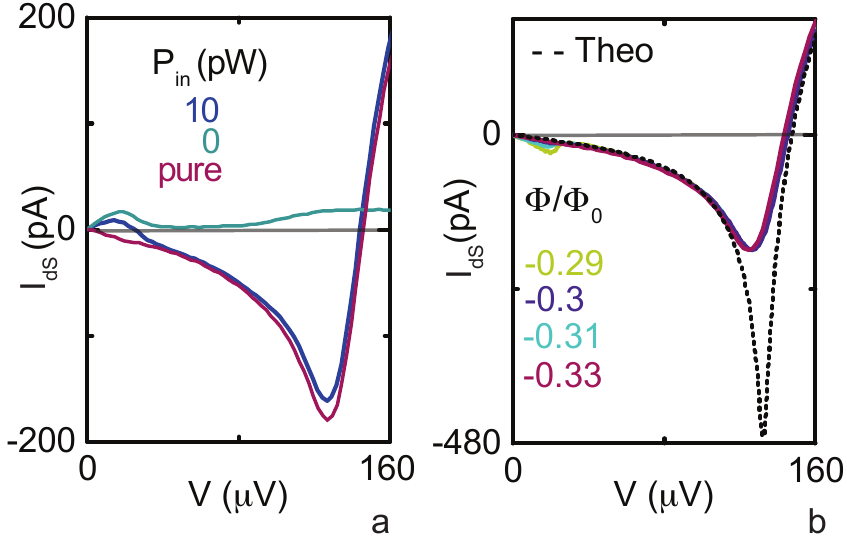}
    \caption{ a) The pure quasiparticle current (red line) extracted from the difference between the current in the presence of a small heating ($P_{in}=$10 pW, blue line) and the current in the absence of a thermal gradient (aquamarine line, $P_{in}=$ 0) at the maximum suppression of the critical current are shown. b) Pure quasiparticle currents for different values of the magnetic flux are compared. The black dashed line represents the fit.}
    \label{Fig7}
\end{figure}

In this section, we explore how the Josephson coupling impacts on thermoelectricity.
In our device, the size of the Josephson current can be controlled by an external magnetic flux thanks to the double-loop geometry (see Secs.~\ref{Sec:Fab&SetUp}-~\ref{sec:SuperEq}).\\
Figure~\ref{Fig6}a displays the $I_{dS}V$ characteristics for different values of $\Phi$, fixing the injection power where the thermoelectric effect is maximum ($P_{in}=$ 10 pW). By increasing the Josephson coupling with the magnetic flux (in the range $\Phi/\Phi_0=[-0.33,-0.24]$), the absolute value of the thermoelectric current $|I_{\rm Th}|$ decreases until it disappears. Instead, $V_{\rm Th}$ turns out to be less sensitive to a variation of $\Phi$ until thermoelectricity is completely suppressed. These behaviors are the results of the competition between Cooper pairs and quasiparticle tunneling. Indeed, when the dissipationless Cooper pairs prevail on the quasiparticle transport, thermoelectric phenomena are non-visible. This interplay is confirmed in Fig.~\ref{Fig6}b, where the absolute value of the thermoelectric current at the peak [$I_{Th}(V_p)$, blue squares] and the maximum of the critical current ($I_{c}$, golden squares) are shown as a function of $\Phi$. These two currents display opposite trends: the thermoelectric current is maximum when the Josephson current is minimum. In particular, the thermoelectric behavior ($|I_{Th}|>0$) starts manifesting for $I_c\lesssim 800$~pA, as highlighted by Fig.~\ref{Fig6}b. Therefore, thermoelectricity occurs when the critical current is roughly $20\%$ of its maximum value, in reasonable agreement with theoretical predictions for an aluminum-based tunnel structure~\cite{MarchegianiPRR}.\\
Moreover, we can observe a linear dependence of the thermoelectric current on the supercurrent, as shown in Fig.~\ref{Fig6}c. Indeed, by increasing the Josephson coupling the dissipationless current rises the background of the total current (Eq. \ref{eq:IVphiJJ}). The latter adds to the negative thermoelectric contribution with the consequent linear reduction of the thermoelectric peak. We compare the experimental data extracted from the device so far investigated ($D_1$, colored squares) with a secondary device ($D_2$, violet circles) in Fig. \ref{Fig6}c. 
In summary, the presence of the Josephson current is detrimental for thermoelectricity, since even small values of the supercurrent can effectively shunt the junctions thereby preventing the manifestation of the thermoelectric current.
These evidences may suggest a relation between the two phenomena. The Josephson coupling indirectly influences the magnitude of the thermoelectric effect, and a relevant Josephson contribution can totally hide thermoelectricity~\cite{MarchegianiPRR}. In reality, the situation is more complex.\\
To prove the distinction between the two transport channels and the independence of the quasiparticle component from $\Phi$, we can extract $I_{qp}(V_J)$ in the subgap regime by subtracting the Josephson contribution $I_{j}(V_J,\Phi)$ to the thermoelectric curves $I_{dS}(V_J,\Phi)$~\footnote{For simplicity, here we identify $V\sim V_J$, being the voltage drop in the filters is quite low in the subgap-regime.}.
Indeed, the measured currents can be considered as the direct sum of the Josephson and the quasiparticle contributions 
$I_{dS}(V_J, T_1, T_2,\Phi) =I_J(V_J,T_1, T_2,\Phi) + I_{qp}(V_J,T_1, T_2)$, where we made explicit the dependence on the electronic temperature $T_i$ relevant in the out-of-equilibrium regime ($T_1 \neq T_2$ for $P_{in}\neq 0$).
In the presence of thermoelectricity, the determination of the Josephson contribution is not straightforward.\\
For our goal, we assume a Josephson contribution weakly dependent on the thermal gradient ($T_1>T_2> T_b$) across the d-SQUID, i.e., $I_J(V_J, T_1, T_2, \Phi)\approx I_J(V_J, \Phi)|_{T_b}$ for small $P_{\rm in}$. Differently, the quasiparticle current is significantly affected by $P_{\rm in}$, and its sign changes in the presence of thermoelectricity.
At thermal equilibrium ($T_1=T_2=T_b$ for $P_{in}=$ 0), we expect the charge current to be dominated by the Josephson contribution in the subgap region [$V_J\ll (\Delta_1 (T_b)+\Delta_2(T_b))/e$], since the dissipative quasiparticle current is exponentially suppressed as $\propto e^{-\Delta_{0,1}/(k_B T_b)}$ for $k_B T_b\ll \Delta_1(T_b)$. Thus,
we approximate the Josephson contribution with the total charge current measured at $T_b=$ 30 mK, i.e.,  $I_J(V_J,\Phi)|_{T_b}\approx I_{dS} (V_J,\Phi)|_{T_b}$.
In this scenario, the pure quasiparticle current, dependent on $V_J$ and $\Phi$, is extracted as follows: $I_{qp} (V_J,T_1, T_2,\Phi)\approx I_{dS}(V_J,T_1,T_2,\Phi)-I_{dS}(V_J,\Phi)|_{T_b}$.
As an example, the extracted pure quasiparticle current (purple line) at the base temperature and for $\Phi/\Phi_0=-0.33$ is shown in Fig.~\ref{Fig7}a. For a comparison, we display as well the measured $I_{dS}V$ characteristics used in the extraction: the subgap charge currents for $P_{in}=$ 10 pW (blue line) and $P_{in}=$ 0 (aquamarine line). 
In the pure quasiparticle curve, we note that around zero bias the Josephson effect is completely removed, and the ANC appears. This result is in good agreement with the spontaneous PH symmetry breaking theory ~\cite{MarchegianiPRL}. Moreover, the absolute value of the thermoelectric current for $V_J\sim V_p$ results slightly higher than the measured value.\\
Our procedure is based on reasonable assumptions in the different regimes, and can be validated by comparing the $I_{dS}V$ characteristics for different $\Phi$.
Indeed, we observe a very good agreement between the quasiparticle currents extracted for different values of $\Phi$, as shown in Fig.~\ref{Fig7}b . This result highlights that the magnitude of the Josephson contribution is modulated by the magnetic flux, but the latter does not impact on the quasiparticle contribution. Thus, thermoelectricity is not affected by the flux variation, being the quasiparticle current dependent only on the DoS of the superconducting leads \cite{MarchegianiPRL}.\\
Notably, the fitting curve is in overall good agreement with the experimental data, even above the singularity-matching peak values, which are not included in the fit. By contrast, a difference in the amplitude of the thermoelectric current  appears in a narrow region around the singularity-matching peak, where the renormalization of the divergence predicted by the BCS theory might not be accurately captured by the simple phenomenological Dynes parameter. 
Furthermore, another possible explanation for the above difference can lie in the deviation of DoS of the Al/Cu bilayer forming the interferometer from the standard  BCS result \cite{FominovFeigelmanPRB63,CatelaniPRB97}.\\


\subsection{Thermoelectric behavior at different bath temperatures}
\label{Sec:TEBath}
\begin{figure}[t!]
    \centering
    \includegraphics[width=1 \columnwidth]{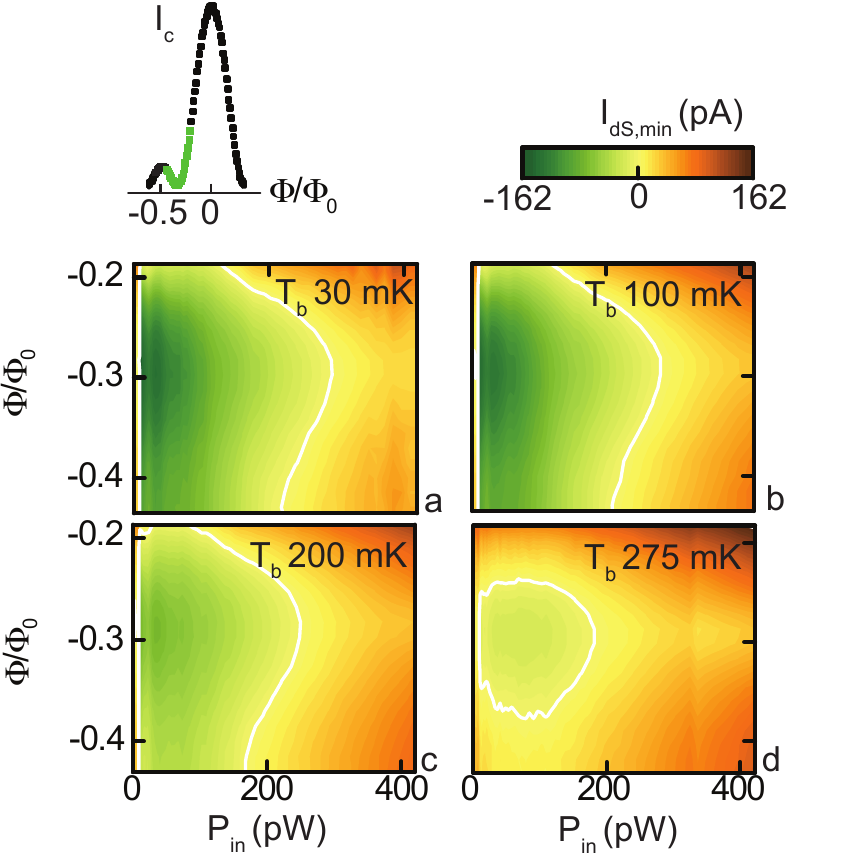}
    \caption{Color maps of the minimum current $I_{dS,min}$ for $V>0$ 
    as a function of the injected power and normalized magnetic flux at a) $T_b=$ 30 mK b) $T_b=$ 100 mK c) $T_b=$ 200 mK d) $T_b=$ 275 mK. Green zones refer to the generated thermoelectric current, whereas orange zones to the dissipative current. The white solid lines provide the zero-current profile ($I_{dS,min}=0$) separating the thermoelectric and dissipative regimes. Top left panel emphasizes the critical current $I_c$ values corresponding to the considered magnetic flux interval in the color maps (green  part of the curve.)}
    \label{Fig8}
\end{figure}

In this section, we shall comment on the thermoelectric current behaviour as a function of the injected power and the magnetic flux for different values of the bath temperature.\\
Figure~\ref{Fig8} displays the minimum current measured ($I_{dS,min}$) for positive voltage bias ($V>0$) as a function of $P_{in}$ and $\Phi/\Phi_0$ for four different bath temperatures (panels a-d). The magnetic flux values are restricted around the first minimum of the interferometer pattern for $\Phi<$ 0 (top-left panel of Fig.~$\ref{Fig8}$). The green areas refer to the thermoelectric current (negative current, i.e., $I_{dS}V<0$), while the orange one to the dissipative current (positive current, i.e., $I_{dS}V>0$). The two different regimes are separated by a white solid line that represents $I_{dS}=0$ detected as the minimum of $I_{dS}V$ characteristics.
In the ranges considered, we can appreciate a sizable variation of the magnitude of thermoelectricity by changing $P_{in}$, while we observe a smooth change of it by varying $\Phi$. The maximum thermocurrent is generated around $\Phi/\Phi_0 \sim $-0.3, which corresponds to the maximum suppression of the Josephson current.
The slight shift registered in $\Phi$-axis with respect to the values expressed so far is due to possible fluxons trapping.\\
The magnitude of the thermoelectric current, expressed as $|I_{\rm dS,min}|$, lowers by increasing $P_{in}$ till its full suppression for $P_{in}>400$~pW. Taking into account the variation along the $P_{in}$-axis, we can note a smaller thermoelectric current around $\Phi/\Phi_0\sim$-0.2, where the Josephson contribution is not so damped.\\
Comparing the four panels of the Fig.~$\ref{Fig8}$, we observe that the thermoelectric behaviour is weakly dependent on $T_b<$ 150 mK. Then, a substantial reduction of the effect is remarked by increasing the bath temperature until $T_b=$ 275 mK, above which thermoelectricity vanishes completely ($T_b=$ 325 mK).
The reason lies on heating of the two superconductors and a consequent changing of the thermal condition across the tunnel junctions, which leads to the suppression of thermoelectricity in the device.
Moreover, we can note the shift of the maximum of the generated current towards larger $P_{in}$ by increasing $T_b$. The reason probably lies on the need of the system to require more input power to establish an adequate thermal gradient between $S_1$ and $S_2$ to generate thermocurrent at higher bath temperature. In this context, if we consider the steady-state thermal balance of the system, it reads $P_{in} = P_{e-ph,S_1} + P_{S_1-S_2}+ P_{S_1}$, where $P_{e-ph,S_1}$ represents the power loss due to the electron-phonon coupling in $S_1$, while $P_{S_1-S_2}$ and $P_{S_1}$ are the power losses toward $S_2$ and the tunnel probes, respectively. In particular, the $P_{e-ph,S_1}$ term turns out to be exponentially suppressed at low temperature, due to the presence of the energy gap in the superconducting DOS \cite{Timofeev}.
At sufficiently  low temperatures  $T_b$, the electronic temperature of $S_1$ can be somewhat different from that of the lattice (i.e., $T_1> T_{ph,1}=T_b$). Therefore, the electron-phonon thermalization ($P_{e-ph,S_1}$) increases by enhancing the bath temperature and, consequently, a higher input power is necessary in order to generate an appropriate thermal gradient across the structure~\citep{GiazottoReview, HeikkilaReview}.\\
To summarize, two different issues seem to contribute to the disappearance of thermoelectricity: the predominance of the Josephson coupling, which short-circuits the junctions therefore hiding the effect, and the deviation from the suitable thermal conditions required in order to obtain a thermoelectric response of the system.\\
\section{Thermoelectric Engine}
\label{Sec:Engine}
\begin{figure}[t!]
    \centering
    \includegraphics[width=1 \columnwidth]{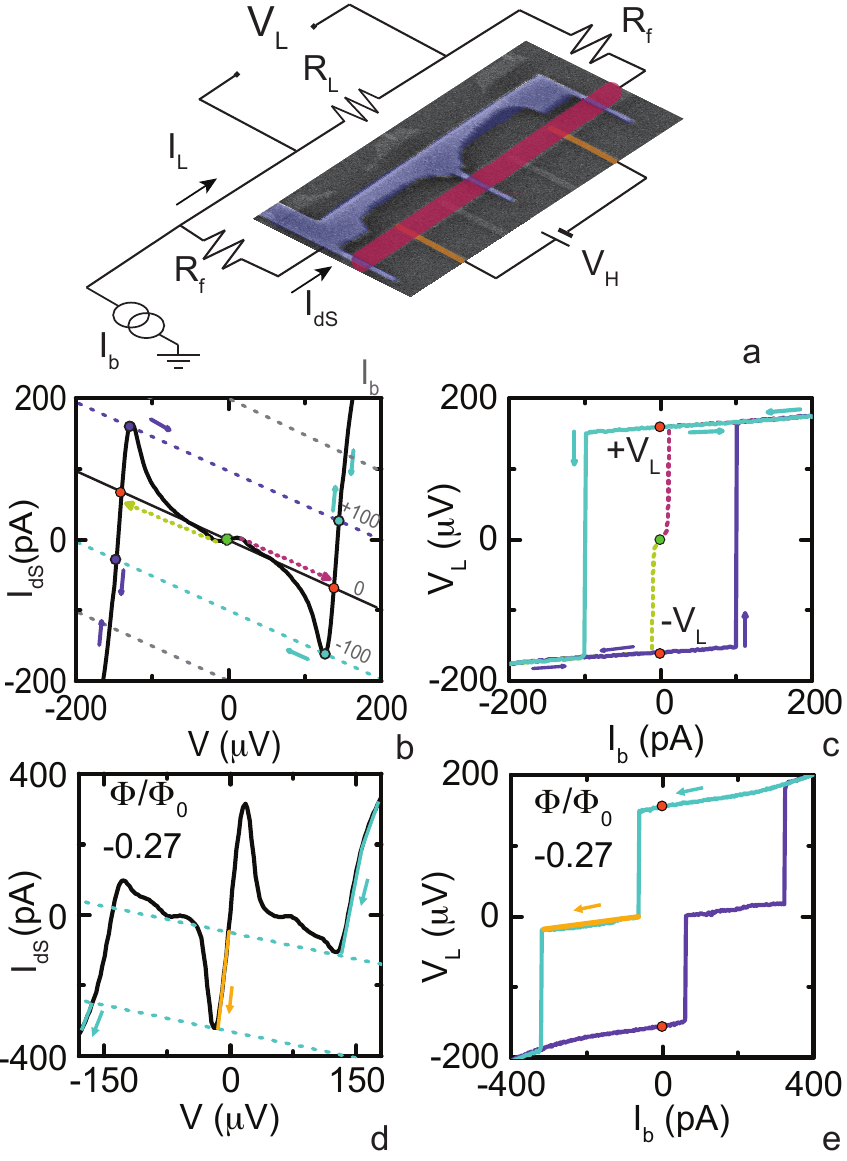}
    \caption{a) Schematic electrical circuit of the parallel connection between the thermoelectric element (SEM picture) and the load resistor ($R_L$), supplied with a current generator ($I_b$). The thermal gradient across the thermoelectric element is guaranteed by a floating voltage source ($V_{H}$). The output voltage generated ($V_L$) is measured across the load resistance. $I_L$ and $I_{dS}$ are the currents that flow in $R_L$ and the d-SQUID, respectively. b) The $I_{dS}V$ characteristic (at $P_{in}=$ 10 pW and $\Phi/\Phi_0=$0.33) is crossed by load-lines with $R_L=$ 2 M$\Omega$ by changing the bias current (the intercept with $I_{dS}$-axis). The colored circles are the intersections between the load line and the $I_{dS}V$ characteristic of the device, yielding the solutions of the Eq.~\eqref{Eq:rettaDicarico}. The red and yellow arrows correspond to the switch-on of the heat engine. On the right, the $I_b$ values corresponding to the intercept with $I_{dS}$-axis are reported in gray. c) The hysteretic behavior obtained from the output voltage is shown as a function of the bias current. The red and yellow curves correspond to the switch-on of the heat engine starting from the $(I_b, V_L)=(0, 0)$ (green circle). The voltage values generated in the absence of the critical current $I_b=0$ are $\pm V_L$ (red circles). d) The $I_{dS}V$ characteristic where the critical current is not strongly suppressed (with $P_{in}=$ 10 pW and $\Phi/\Phi_0=$ -0.27) is shown. The orange line corresponds to the solutions given by the intersections of the load-lines with metastable state determined by the Josephson contribution. e) The hysteretic cycle presents a step around $V_L=$0 due to the supercurrent that flows in the d-SQUID for those values of the injected bias current.
    }
    \label{Fig9}
\end{figure}
In this section, we discuss a concrete application of the thermoelectric element as a heat engine. We show an intriguing consequence of the spontaneous breaking of the PH symmetry in the device, which induces an hysteretic behaviour. In particular, we investigate the influence of the Josephson coupling on the engine operation.\\ 
For this purpose, we connect in parallel the d-SQUID with a load resistor ($R_L$), and we supply the whole system with an external current source ($I_b$) (see the schematic electric circuit displayed in Fig.~\ref{Fig9}a). In this circuit,
the thermoelectric element is in series with the two RC filters resistances ($R_f$) of the cryostat measurement lines, and $V$ is the total sum of the voltage drop across them. Our device is properly energized through two heating probes by a floating circuit ($V_H$) to ensure a suitable thermal gradient between the two superconductors, as previously discussed. 
The output voltage ($V_L$) is then measured across the load resistance. The electrical circuit can be resolved by the following system of equations:
\begin{equation}
\label{Eq:sistema}
\begin{cases}
    I_{b}=I_{L}+I_{dS}(V) \\
    I_{L}=\frac{V_L}{R_L} \\
    \end{cases}
\end{equation}
where $I_L$ and $I_{dS}(V)$ are the currents flowing through the load and the d-SQUID, respectively.
Being a parallel circuit, the voltage across the resistance ($V_L$) is equal to the sum ($V$) of voltage drops across the RC-filters and the thermoelectric element. Therefore, we can derive the relation of the current flowing in the device as:
\begin{equation}
   I_{dS}(V) = -\frac{V}{R_L} + I_{b}.
    \label{Eq:rettaDicarico}
\end{equation}
The current $I_{dS}$(V) depends on the voltage developed across the parallel $V$, and coincides with the $I_{dS}V$ characteristic of the d-SQUID previously investigated (see Fig.~\ref{Fig1}b). The load resistance determines the slope of the load-line ($-1/R_L$), and the external current source $I_b$ fixes its intersections with the $I_{dS}$-axis (see Fig.~\ref{Fig1}b).
We can thus determine self-consistently the solutions of Eq.~\ref{Eq:rettaDicarico} as a function of the current bias $I_b$.
As shown in Fig.~\ref{Fig9}b, resolving graphically this relation means to find the intersections (colored circles) between the $I_{dS}V$ characteristic and the load-lines for each value of $I_b$. The solutions ($V$) in Fig.~\ref{Fig9}b correspond to the output voltage ($V_L$) measured and reported in Fig.~\ref{Fig9}c.
It is worthwhile to note that the equation can admit multiple solutions due to the non-monotonic behavior of the thermoelectric curve. The system sets the meta-stable solutions with \textit{positive} differential conductance, $dI_{dS}/dV>0$; the other ones are electrically \textit{unstable}~\cite{MarchegianiPRB}.\\
As a starting point in Fig.~\ref{Fig9}b, we take into account a load resistor of $R_L=$ 2 M$\Omega$ ($I_b=0$, black solid line), which crosses at zero-current bias the $I_{dS}V$ characteristic, measured at  $P_{in}=$ 10 pW and $\Phi/\Phi_0=$ -0.33. By changing the current bias the load-line sweeps the whole thermoelectric characteristic (dashed colored lines). Linking all the intersections, we can obtain the hysteresis curve as the one measured and shown in Fig.~\ref{Fig9}c.\\ 
At the beginning, for $I_b=$ 0, the system stays in $(I_b,V_L)=(0,0)$ in a metastable state (see Fig.~\ref{Fig9}c), which corresponds to the central green solution in Fig.~\ref{Fig9}b.
This metastable state is determined by a small residual Josephson contribution, which traps the system around $V\approx 0$.
By injecting a positive (negative) current bias, the system is moved through red positive (yellow negative) voltage values (see Fig.~\ref{Fig9}c), crossing the residual dissipationless Josephson current in Fig.~\ref{Fig9}b. Continuing to increase the absolute value of $I_b$ in Fig.~\ref{Fig9}c, the system jumps on the thermoelectric region and moves towards the dissipative region of the $I_{dS}V$ characteristic. On the contrary, by decreasing $I_b$ the system remains trapped in the thermoelectric regime, and it is able to generate a non-zero output voltage $\pm V_L(I_b=0)$ even in the absence of an external bias current, which corresponds to the red circle solutions appearing in Fig.~\ref{Fig9}b-c. The thermoelectric generation is sustained even when $I_b$ changes sign, until it reaches the thermoelectric peak at $I_b=$ -100 pA. By further increasing $|I_b|$ the devices cannot support anymore thermoelectricity, and the system switches to the other dissipative branch of the hysteresis loop (see Fig.~\ref{Fig9}c), which corresponds to the dissipative side of the $I_{dS}V$ characteristic at $V<0$ (Fig.~\ref{Fig9}b). In the same way, the other branch of the hysteresis curve is obtained by reducing (in module) the current bias back to zero.\\
For a sizeable Josephson current contribution (i.e., for $I_c= 325$~pA at $\Phi/\Phi_0=-0.27$), the situation is qualitatively different (see Fig.~\ref{Fig9}d). 
After sweeping the $I_{dS}V$ branch for positive voltage until to the thermoelectric peak (aquamarine line in Fig.~\ref{Fig9}d), the load-line crosses the negative peak of the Josephson contribution around $V\sim 0$ (orange line in Fig.~\ref{Fig9}d). These solutions correspond to the orange step around $V_L=0$ in the hysteresis cycle (Fig.~\ref{Fig9}e). The higher the Josephson contribution the larger  is the stability range  around $V\sim 0$ in the hysteresis loop. 
The system continues to be able to sustain two metastable states ($\pm V_L$) at $I_b=$ 0 also in the presence of a sizeable Josephson contribution underlying once again the bipolarity of the effect.\\
These metastable states can take the role of voltage levels of a thermoelectric memory element controlled by a bias current ~\cite{MemoriaBrevetto}.


\subsection{Phase-dependent engine performance}
\begin{figure}[t!]
    \centering
    \includegraphics[width=1 \columnwidth]{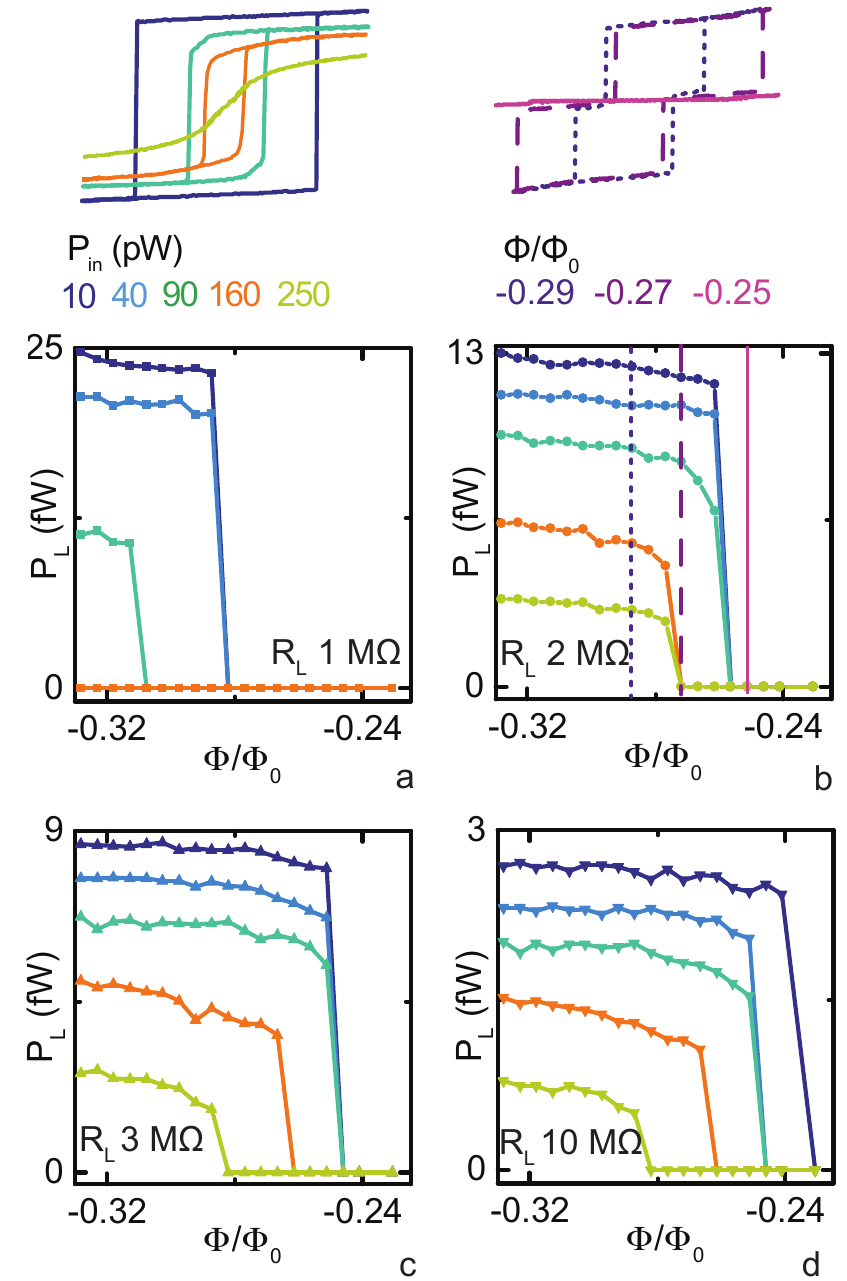}
    \caption{The output power delivered to the load resistor as a function of the magnetic flux for different values of the injected power and zero current bias. a) $R_L=$ 1 M$\Omega$, b)$R_L=$2 M$\Omega$, c) $R_L=$3M$\Omega$, d) $R_L=$ 10 M$\Omega$. All data are extracted at $T_b=$ 30 mK. On the top: left side, the hysteresis loop by changing the input power; right side, the hysteretic behaviour by changing the magnetic flux.}
    \label{Fig10}
\end{figure}
Let us now focus on the output power generated by the thermoelectric engine without any external bias current  ($I_b=0$) by tuning the Josephson contribution. We fix the values of the heating power, and compare the output generated for different load resistances.\\
In the previous section, we introduced the explanation of the hysteresis curves in the absence and the presence of the Josephson contribution. 
In the first case, i.e., working where the supercurrent is maximally suppressed, we can observe a reduction of the width and the height of the hysteresis loop by increasing the injected power until we close completely the cycle (see  top left image of Fig.~\ref{Fig10}). Indeed, the thermoelectric element ceases to generate an output power when thermoelectricity disappears. In the second case, moving away from the point of the maximum suppression of the supercurrent, the Josephson contribution increases, and, consequently, the steps around $V_L=0$ enlarge until achieving full suppression of the hysteresis (see top right image of Fig.~\ref{Fig10}). As a matter of fact, when the Josephson current in the device is too large, the thermoelectric phenomena are not visible.\\ 
We want now to compare the output power ($P_L=V^2_L/R_L$) generated at $I_b=0$ as a function of three main parameters: the injected heating power, the magnetic flux, and the load resistance. 
By fixing $\Phi$ and $R_L$, we note that $P_L$ decreases by increasing the input power in all four panels of Fig. $\ref{Fig10}$. These results are consistent with the intensity reduction of the thermoelectric peak, and its shift towards lower voltage. The maximum output power ($P_L=25$~fW) is obtained for $P_{in}=$ 10 pW, and, clearly, for the minimum resistance allowed.\\ 
On the other hand, by fixing $P_{in}$ and $R_L$, the output power is almost constant by varying $\Phi$, until the supercurrent starts to screen the thermoelectric generation, as predicted in Ref.~\cite{MarchegianiPRR}. This behavior remarks the weak dependence of the thermovoltage [$V_{Th}\equiv V_L (I_b=0)$] on $\Phi$, until thermoelectricity is displayed. Indeed, as previously observed, tuning the supercurrent leads to a sizeable variation of the thermoelectric peak intensity, but weakly affects the position of the Seebeck voltage. Therefore, both the thermovoltage and the generated power  remain almost constant.\\
Lastly, we wish to comment on how $P_L$ changes by changing $R_L$. By increasing the load resistor (see from Fig.~\ref{Fig10}a to Fig.~\ref{Fig10}d), the range of  magnetic flux and injected power where  power generation is present are ever wider. The reason lies in the smaller slope of the load line ($-1/R_L$) that permits crossing the thermoelectric peak even though it is significantly reduced by an unfavorable thermal condition, or screened by a sizable Josephson contribution.
Thus, the slope of the load line is crucial for the operation of the thermoelectric engine. In particular, for small values of $R_L$ the load line never intercepts the thermoelectric peak. Indeed, the minimum value of the load resistance that can be supported by the junction is $R_{L,min}=V_{p}/I_{Th}(V_p)= 0.8$ M$\Omega$, and  no power can be generated if $R_L<R_{L,min}$.\\

\section{Impact of inhomogeneity of multiple junctions on the thermoelectric response}
\label{Sec:Non_idealities}
As broadly discussed in Sec.~\ref{sec:SuperEq}, the double-loop SQUID geometry of our device is used to maximally reduce the Josephson coupling even in the presence of unwanted asymmetries between the junctions. Hereinafter, we investigate the role of inhomogeneities of the tunnel junctions that could potentially affect the thermoelectric behavior of the interferometer.\\
In our device, the voltage drop across the three parallel $S_1IS_2$ tunnel junctions is the same, while the currents flowing through each of them are inversely proportional to the tunnel resistance. Thus, the parallel connection enhances the generated thermoelectric current resulting from the sum of each junction contribution. \\
Here, we theoretically discuss a hypothetic device composed by $n$ different $S_1IS_2$ junctions labeled with the index $l=\{1,\dots n\}$. For the sake of clarity, we assume that they are totally independent from each other, despite what happens in our device where the three tunnel junctions share the same superconductors. The specific $l$th-junction is characterized by a normal-state resistance $R_{T}^{(l)}$, a zero-temperature energy gap $\Delta_{0,j}^{(l)}$ and an electronic temperature $T_{j}^{(l)}$, where $j=$ 1, 2 is the lead index within each junction.
Thus, the total current in the parallel configuration can be written as
\begin{equation}
I_{tot}=\sum_l I_{qp}(R_{T}^{(l)},\Delta_{0,1}^{(l)},\Delta_{0,2}^{(l)},T_1^{(l)},T_2^{(l)},V_J),
\label{Eq:sommatoria}
\end{equation}
where the expression for $I_{qp}$ is given by Eq.~\eqref{eq:iqp}. For simplicity, we analytically address the dependence of the current on the variation of a single parameter.
In the simplest case, we can consider only the variation of the tunnel resistance ($I_{qp} \propto 1/R_{T,i}^{(l)}$), concluding that $I_{tot}$ is proportional to the parallel resistance of all junctions $1/R_{T,tot}=\sum_l 1/R_{T}^{(l)}$.\\
A different approach is required when we take into account another parameter $Y$ different from the resistance. For $n\gg 1$, we can replace the summation in Eq.~\eqref{Eq:sommatoria} with an integral over the inhomogeneous parameter $Y$
\begin{equation}
I_{tot}=\int dY P(Y) I_{qp}(Y,V),
\end{equation}
where $P(Y)$ is the distribution function, and $I_{qp}(Y,V)$ is the current through a single junction.
For a large number of junctions, the probability distribution takes a Gaussian form due to the central limit theorem, i.e., $P(Y)=(\sqrt{2\pi\sigma_Y^2})^{-1} \exp[-(Y-\bar{Y})^2/2\sigma_Y^2]$, with $\bar{Y}$, $\sigma_Y^2$ the mean and variance of the stochastic variable, respectively.
When two or more parameters fluctuate simultaneously, the case can be generalized to a multivariate distribution by replacing $Y\to \vec{Y}$, where $\vec{Y}=(\Delta_{0,1},..,T_2)$ represents the vector of all varying parameters of the junction. Consequently, $P(\vec{Y})$ becomes the probability distribution. When the parameters present a very small variation, they can be treated as independent variables $P(\vec{Y})\approx P(\Delta_{0,1})...P(T_2)$. In general, a small Gaussian distribution of the fluctuating junction parameter is expected to smooth the $I_{dS}V$ characteristics.\\
In order to discuss the effect of the junctions inhomogeneity on the $I_{dS}V$ characteristics, it is convenient to simplify the system by considering only two junctions.
\begin{figure}[t!]
    \centering
    \includegraphics[width=1 \columnwidth]{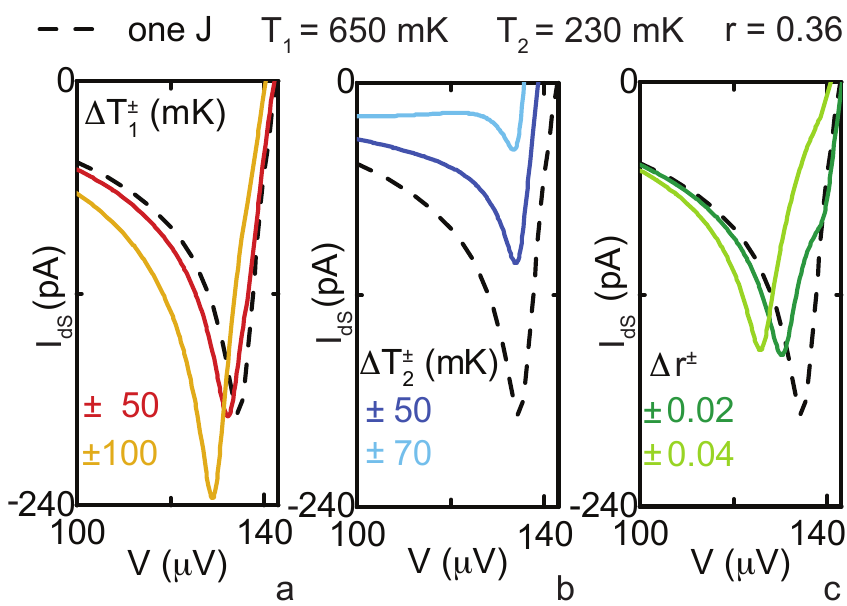}
    \caption{The comparison of the thermoelectric peaks are shown in the subgap regime for positive voltage at the minimum thermal gradient required. The current generation obtained simulating a single Josephson junction is reported in black dashed line ($T_1= 650$ mK, $T_2= 230$ mK and $r=$ 0.36). The colored solid lines refer to the parallel between two $S_1IS_2$ tunnel junctions simulating: a) a thermal difference between the two hot leads ($\Delta T_1^\pm=\pm$ 50 mK, red line and $\Delta T_1^\pm=\pm$ 100 mK, orange line), b) a thermal difference between the two cold leads ($\Delta T_2^\pm=\pm$ 50 mK, blue line and $\Delta T_2^\pm=\pm$ 70 mK, light blue line), c) a difference between the superconducting energy gaps in the two leads ($\Delta r^\pm=\pm$ 0.02, green line and $\Delta r^\pm=\pm$ 0.04,  light green line). The average values of $T_1$ and $T_2$ are extracted from the thermometers calibration, and the $r$ value from the experimental data.}
    \label{Fig11}
\end{figure}
Therefore, we adapt Eq.~\ref{Eq:sommatoria} by fixing the phenomenological Dynes parameter for the two superconductors at the values previously found, and assuming a BCS DoS for the superconducting bilayer.\\
We start by considering first of all only the variation of the temperature of the hot lead ($T_{1}^{(l)}$ with $l=\pm$ the junction index). Figure~\ref{Fig11}a reports the comparison between three thermoelectric curves. We simulate a single junction with $T_1=$ 650 mK (black dashed line) and two examples with two parallel junctions, which have a temperature of $T_{1}^{(\pm)}=T_1 \pm$ 50 mK (red line) and  $\pm$ 100 mK (orange line), respectively. The shift of the thermoelectric peak is due to the expected change in the BCS gap $\Delta_{1,2}(T_{1,2})$ dependent on the lead temperature variation.\\
Similarly, we can consider a variation of the temperature of the cold lead. In Fig.~\ref{Fig11}b, we show the thermoelectric curves assuming the temperature of the two junctions to be  $T_{2}^{(\pm)}=T_2 \pm$ 50 mK (blu line) in the first case, and $\pm$ 70 mK (light blu line) in the second case, respectively. The mean value considered is $T_2=$ 230 mK, which corresponds to the cold temperature lead for a single junction (black dashed line). We can observe a reduction of the thermoelectric peak driving the total current of the device in the dissipative regime, when the cold lead temperature is too high. In this case, the thermopower can be totally suppressed, as predicted also in Ref.~\cite{MarchegianiPRB}.\\
Instead, in Fig.~\ref{Fig11}c we show what it does happen by changing the zero-temperature energy gap of the two hot leads. We suppose a variation of the gaps ratio of $r^{(\pm)}=r \pm$ 0.02 (green line), and $r \pm$ 0.04 (light green line) with respect to the mean value of 0.36 for a single junction.
In this last case, we observe an additional bump appearing on top of the main thermoelectric peak. This situation was shown to be present also in Ref.~\cite{GermaneseArxiv}, where for high input powers some $I_{dS}V$ characteristics presented such an additional shoulder. This feature present in the experiment might thus be ascribed to a gap inhomogeneity between the junctions composing the interferometer.\\ 
Inhomogeneity in the junction parameters could partially account for the mismatch between the experimental data and the theoretical modeling for $V\sim V_p$ (Fig.~\ref{Fig4}b).\\
It is important to emphasize that a similar analysis can be carried out as well with tunnel junctions connected in series, where thermoelectricity can still occur due to the intrinsic bipolarity of the phenomenon, and current conservation~\cite{MarchegianiPRB,GermaneseArxiv}. In such a case, the total voltage drop occurring across the junctions series is the sum of all voltage drops occurring across each single junction. Since the current flowing in each junction is not monotonic in the voltage bias, a more complex nonlinear analysis is required to solve the problem. Therefore, a full numerical solution is not shown here, since it is not so relevant for our experiment.\\ 
In conclusion, we have shown that the presence of  inhomogeneities in the parameters characterizing  
the different junctions composing the system may  affect the thermoelectric response of the whole device.

\section{Conclusions}
As reported in Ref.~\cite{GermaneseArxiv}, we have demonstrated that in the presence of a large thermal gradient, thermoelectricity is generated in a structure composed by three $S_1IS_2$ Josephson tunnel junctions arranged in a parallel configuration and forming a double-loop superconducting quantum interference device. 
The bipolar thermoelectric effects here shown are the result of a spontaneous particle-hole symmetry breaking, which is predicted to occur in a superconducting tunnel junction where two superconductors with different energy gaps are kept at different temperatures \citep{MarchegianiPRL, MarchegianiPRB}. 
Our interferometer setup has allowed to  suppress  the Josephson coupling present in the structure up to a large extent, thereby getting a maximum thermovoltage of $\sim\pm$ 150 $\mu$V for $\delta$T $\sim$ 400 mK. In addition, when connected in parallel to a generic load resistor, the bipolar thermoelectric engine can deliver an electric output  power up to $\sim25$ fW, which corresponds to a sizable power-density of 140 nW/mm$^2$.\\
Moreover, the phase-tunability of the thermoelectric effect allows a phase-coherent control of the engine performance. In particular, the thermoelectric current and output power can be tuned by changing the magnetic flux piercing the interferometer loops. 
Here, we have investigated the impact of the Josephson contribution on thermoelectricity. Notably, we observe a robust effect likewise.
The thermoelectric phenomena associated to the quasiparticles transport are indirectly influenced by the Josephson contribution, which completely short-circuits the junctions when it is too large. 
In particular, we have demonstrated the phase-independence of the thermoelectric effects by removing the Josephson contribution from the measured thermoelectric current-voltage characteristic. 
This result underlines that the thermoelectric production in a DC system is to be ascribed only to the quasiparticle term determined by the phase-independent DoS of the two leads. 
In the same way, we have proved the engine operation by varying the Josephson coupling for fixed values of the temperature difference between the two superconductors. In particular, we have shown that the Josephson contribution introduces an extra metastable state at $V\approx 0$, affecting the hysteretic loop and extending the applicability domain of the device.\\
At the end, possible sources of inhomogeneity due to structural differences between the three tunnel junctions are discussed. Thermal gradients, and different zero-temperature energy gaps have been considered.
Homogeneous heating along the $S_1$ strip is not so obvious. A thermal gradient may be created across the heated superconductor influencing differently the three junctions at the overlap, which could show different tunnel resistances. 
As a consequence, also the cold leads may suffer from this thermal inhomogeneity since they are not thermally isolated. It is not simple to identify which parameters are the main cause of the small discrepancy between the experimental data and the theoretical prediction, thus further investigations have to be carried out.\\
Finally, a possible implementation of the bipolar thermoelectric Josephson engine is envisioned, which exploits the parallel connection of $n$ thermoelectric elements so to increase the total generated thermocurrent and output power.\\
Our prototypical superconducting thermoelectric device might find direct application in quantum technology \citep{Ladd, Siddiqi, PoliniArxiv} through the implementation of engines, power generators, electronic devices \cite{Braginski}, memories \cite{MemoriaBrevetto}, radiation
sensors \cite{Heikkila2018}, and switches. The presence of the Josephson contribution could be used as an on/off switch for thermoelectricity in such tunnel junctions, thereby allowing a phase-coherent control of the system at subKelvin temperatures. 
Yet, the above described bipolar thermoelectric effect is expected to occur as well in several thermally-biased physical systems, which are
characterized by an intrinsic PH symmetry, and  where the hot and cold electrodes possess a gapped and monotonically-decreasing density of states, respectively~\cite{MarchegianiPRL}. 
For this reason, our study is pivotal for groundbreaking investigations of nonlinear thermoelectric
effects in a number of different solid-state systems ranging from semiconductors and low-dimensional electronic materials\cite{BernazzaniArxiv} to high-temperature
superconductors and topological insulators.

\section*{Acknowledgements}
The authors wish to acknowledge the EU’s Horizon 2020 research and innovation program under Grant Agreement No. 800923 (SUPERTED) and No. 964398 (SUPERGATE) for partial financial support. A.B. acknowledges the Royal Society through the International Exchanges between the UK and Italy (Grants No. IEC R2 192166 and IEC R2 212041)

\bibliographystyle{apsrev4-1}

\end{document}